\documentclass[twocolumn,showpacs,preprintnumbers,amsmath,amssymb]{revtex4}

\usepackage{graphicx}% Include figure files
\usepackage{dcolumn}% Align table columns on decimal point
\usepackage{bm}% bold math

\begin{document}
\title{Mott Transition, Compressibility Divergence and P-T Phase Diagram of Layered Organic Superconductors: An Ultrasonic
Investigation.}
\author{D. Fournier}
\author{M. Poirier}
\author{M. Castonguay}
\author{K. Truong}

\affiliation{Centre de Recherche sur les Propri\'et\'es
\'Electroniques de Mat\'eriaux Avanc\'es and D\'epartement de
Physique, Universit\'e de Sherbrooke, Sherbrooke, Qu\'ebec,
Canada J1K 2R1.}

\date{Sept. 2002}
\begin{abstract}
The phase diagram of the organic superconductor
$\kappa$-(BEDT-TTF)$_2$Cu[N(CN)$_2$Cl has been investigated by
ultrasonic velocity measurements under helium gas pressure.
Different phase transitions were identified trough several
elastic anomalies characterized from isobaric and isothermal
sweeps. Our data reveal two crossover lines that end on the
critical point terminating the first-order Mott transition line.
When the critical point is approached along these lines, we
observe a dramatic softening of the velocity which is consistent
with a diverging compressibility of the electronic degrees of
freedom.

\end{abstract}
 \maketitle


Electronic properties of metals or semiconductors at energies of
the order of room temperature or less can usually be understood
by simple models that take into account a few electronic bands
near the Fermi level. In the absence of broken symmetries, it
suffices to take into account the effects on the bands of small
residual electron-electron and electron-phonon interactions to
explain the observed electronic properties. In the last decade, a
flurry of activity has centered on materials that {\it cannot} be
understood using the above textbook procedure. In these
materials, only one or a few bands should be relevant to
understand electronic properties but the residual interactions
are so strong  that electronic eigenstates can become localized,
leading to a complete breakdown of the band picture. A few
compounds have become prototype materials for the study of these
so-called strong correlation effects: $V_2O_3$ \cite{McWhan},
$Ni(S,Se)_2$ \cite{Matsuura} and the family of organic conductors
$\kappa$-(BEDT-TTF)$_2$X \cite{Kanoda}.

In these compounds, one observes a pressure-induced
finite-temperature first-order phase transition (MI) between an
insulating (I) and a metallic (M) phase. Pressure (hydrostatic or
chemical) increases the bandwidth, reducing the effects of
residual interactions. This first-order phase transition seems to
correspond to the so-called Mott transition, as it has become
understood in the last few years through dynamical mean-field
theory (DMFT) \cite{Georges}.

One of the recent predictions of DMFT, is that the
compressibility of electronic degrees of freedom diverges at the
critical point that terminates the first-order Mott transition
line \cite{Kotliar,Onoda}. Observation of this phenomenon would
help confirm the picture of the Mott transition proposed by DMFT.
In this letter, we present the first ultrasonic study of
$\kappa$-(BEDT-TTF)$_2$Cu[N(CN)$_2]$Cl (denoted as $\kappa$-Cl),
using an hydrostatic helium gas pressure cell. The study,
summarized in the phase diagram shown in figure 1, shows (a) A
very large softening of the sound velocity at the critical point,
corresponding to the predicted divergence of the compressibility
of the electronic degrees of freedom. (b) Two crossover lines -
joining at the critical point - where a similar although smaller
compressibility anomaly is observed. Most remarkably, while the
compressibility anomalies decrease in size as one moves away from
the critical point, the crossover line at high pressure coincides
with the well-known pseudogap features identified in the magnetic
\cite{Mayaffre}, transport \cite{Shushko,Frikach} and elastic
properties of the other compounds of the family corresponding to
higher chemical pressure, X = Cu(NCS)$_2$ and Cu[N(CN)$_2$]Br.
This suggests a common origin to the phenomena. (c) In addition,
we show that much smaller signatures in our ultrasonic velocity
measurements confirm the superconducting (SC) phase boundaries
previously identified mostly on the metallic side of the Mott
transition \cite{Lefebvre}. Curiously, our measurements are not
sensitive to the antiferromagnetic phase boundary on the
insulating side of the Mott transition, although one of the
crossover lines is not far from that phase boundary.

\begin{figure}[htbp]
\begin{center}
\includegraphics[width=8cm]{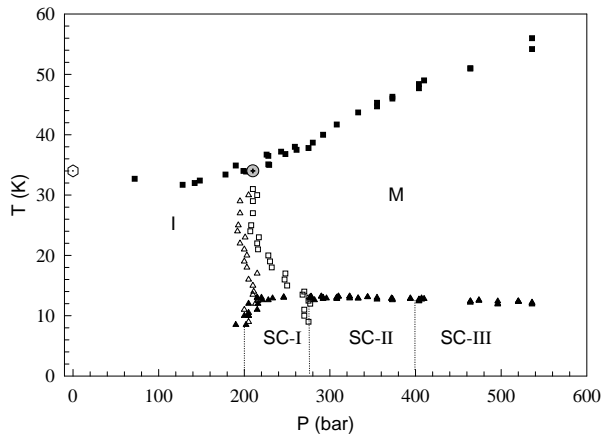}
\vspace{0cm} \caption{\label{}Temperature vs pressure phase
diagram of $\kappa$-Cl. Different symbols are associated to
different anomalies gathered on three different crystals from
temperature sweeps (full symbols) and pressure sweeps (open
symbols). The zero pressure point (dotted hexagon) was obtained
from a microwave resistivity measurement. The gray circle
indicates the critical point ($P_0$, $T_0$).} \vspace{0cm}
\end{center}
\end{figure}

The difficulty to carry out ultrasonic experiments in organic
conductors is primarily due to their very small dimensions
\cite{Frikach}. An additional difficulty in the $\kappa$-Cl
compound comes from small undesirable pressure effects (a crystal
embedded in vacuum grease experiences a pressure of a few
hundreds bars) that can be induced when the piezoelectric
transducers are glued to the crystal faces. Considering the
geometry of single crystals synthesized by the standard
electrocrystallization technique \cite{Wang}, only longitudinal
waves propagating along a direction perpendicular to the layers
(compression mode) could be generated. Because of multiple bonds
and an intrinsically large attenuation at low temperatures, the
measurements could only be done at the fundamental frequency of
the transducer, namely 32 MHz. To ensure hydrostatic pressure
conditions, we restricted the temperature range to the region
above the helium solidification line. Results presented here were
repeated on three crystals having approximate dimensions 1.0 x
1.0 x 0.4 mm$^3$. The points on the phase diagram of Fig.1 come
from these three crystals, so their dispersion represents our
accuracy. Cooling was slow ($\sim$ 0.6 K/min) to avoid any
residual disorder of ethylene end groups. We tested,
simultaneously, the quality of our $\kappa$-Cl crystals at
ambient pressure by measuring the microwave resistivity at low
temperatures \cite{microwave}.

\begin{figure}[htbp]
\begin{center}
\includegraphics[width=8cm]{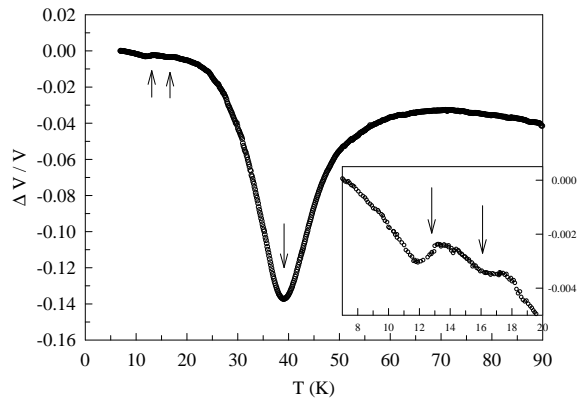}
\vspace{0cm} \caption{\label{}Temperature dependence of $\Delta
V/V$ at 270 bars. The velocity variation is relative to the value
at 7 K. Inset: low temperature portion. ( Arrows indicate the
anomalies).} \vspace{0cm}
\end{center}
\end{figure}

Several elastic anomalies are identified on the relative variation
of the longitudinal ultrasonic velocity ($\Delta V/V$) from
isothermal sweeps. As shown typically in figure 2, three
anomalies can be easily identified at $P = 270$ bars, two small
ones at 13 and 16 K and a huge dip centered at 38 K.

The small anomaly at 13 K characterizes the superconducting
order; this has been confirmed by analyzing the magnetic field
dependence of the anomaly. As previously shown in other compounds
of the family \cite{Frikach}, a small softening anomaly appears in
$\Delta V/V$ below the superconducting critical temperature
$T_c$. The latter is determined by the maximum rate of softening
on decreasing temperature sweeps. We present, in figure 3, low
temperature sweeps of $\Delta V/V$ at selected pressures in the
range 200-500 bars. The pressure range can be divided in three
regions I (200-275 bars), II (275-400 bars) and III (over 400
bars). In region I ($P$ = 245 and 275 bars), the anomaly shows a
significant hysteresis over a wide temperature range above $T_c$;
for increasing temperature sweeps, the anomaly's amplitude is
much smaller than for decreasing sweeps. Moreover, both the
amplitude and $T_c$ increase with pressure. In region II ($P$ =
290 and 325 bars), the hysteresis is much smaller and it merely
represents a tiny shift of $T_c$ between increasing and
decreasing temperature sweeps. In region III ($P$ = 475 bars), no
hysteresis is observed. We have reported in Fig.1 the $T_c$$(P)$
deduced from the temperature sweeps (full triangles). This
$T_c$$(P)$ line is fully consistent with the results of Lefebvre
{\it et al.} \cite{Lefebvre}. Our ultrasound data confirm the
metastability of the SC phase in pressure regions I and II.
Although we cannot specify the nature of the other phase
coexisting with the SC one, NMR and AC susceptibility data have
indicated an inhomogeneous antiferromagnetic phase
\cite{Lefebvre}. Above 400 bars, the finite density condensate
SC-III is characterized by a slowly decreasing $T_c$ with
pressure in agreement with previous results \cite{Lefebvre,Ito}.

\begin{figure}[htbp]
\begin{center}
\includegraphics[width=8cm]{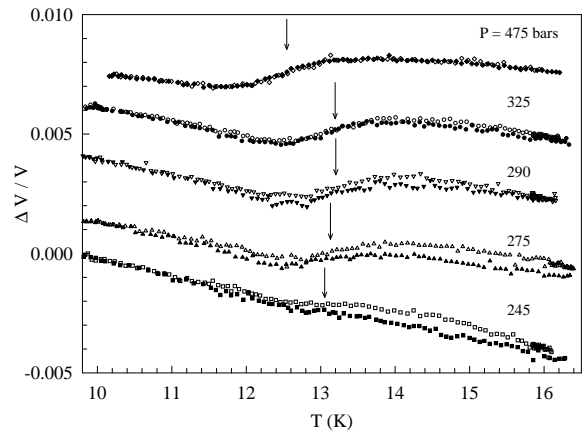}
\vspace{0cm} \caption{\label{}$\Delta V/V$ around 13 K for various
pressures (Full symbols: increasing temperature sweeps; open
symbols: decreasing temperature sweeps). The velocity variation
is relative to the value at 9.8 K. The curves have been shifted
vertically for a better view. The arrows indicate the transition
temperature.} \vspace{0cm}
\end{center}
\end{figure}

We now examine the anomaly in $\Delta V/V$ giving rise to the
huge dip observed at 38 K for $P$ = 270 bars (Fig.2). Its
temperature profile is shown in figure 4 for various pressures.
The amplitude of the dip increases tremendously and shifts to
lower temperatures as the pressure is decreased from 460 bars. It
reaches its maximum amplitude ($> 20\%$) near 34 K and $P$ $\simeq
210$ bars, these values of temperature and pressure defining the
critical point ($P_0$, $T_0$). Below 210 bars, the anomaly is
still observable but its amplitude decreases progressively as the
pressure is lowered to 150 bars, where its position is stabilized
around 32 K.  This huge velocity anomaly, which implies a decrease
of the perpendicular compression elastic constant of almost 40$\%$
at the critical point, is likely due to a strong coupling between
longitudinal acoustic phonons and electronic degrees of freedom.
This points to a compressibility divergence, which has been
described recently as a generic feature of the finite temperature
critical point of the Mott transition \cite{Kotliar,Onoda}. A
similar velocity anomaly has also been observed in compounds X =
Cu(NCS)$_2$ and Cu[N(CN)$_2$]Br having a higher chemical pressure
\cite{Frikach}. We have reported, in Fig.1, the temperature of
the softening dip as a function of pressure (full squares). These
points correspond, at high pressure ($P >$ 210 bars), to the
crossover line between incoherent and coherent metallic phases in
$\kappa$-Cl and X = Cu(NCS)$_2$ and Cu[N(CN)$_2$]Br compounds
(line defined by the peak in $d\rho /dT$ \cite{Shushko}).
Surprisingly, this line does not terminate at the critical point
($P_0$, $T_0$), but continues in the low pressure range reaching
a plateau near 32 K at the lowest pressure attained in our
experiment (75 bars). This line connects smoothly with the
microwave resistivity point at $P$ = 1 bar (dotted hexagon)
\cite{microwave}. These results therefore allow a better
understanding of the transport properties within the PI phase: at
high pressures ($P > P_0$) along the crossover line, $d\rho /dT$
is positive and it shows a peak which moves to higher temperature
with increasing pressure whereas, along the low pressure portion
of the line ($P < P_0$), $d\rho /dT$ is rather negative and shows
a dip \cite{Ito}. Above the critical point, one can then pass
continuously from an incoherent metal to an insulator.

\begin{figure}[htbp]
\begin{center}
\includegraphics[width=8cm]{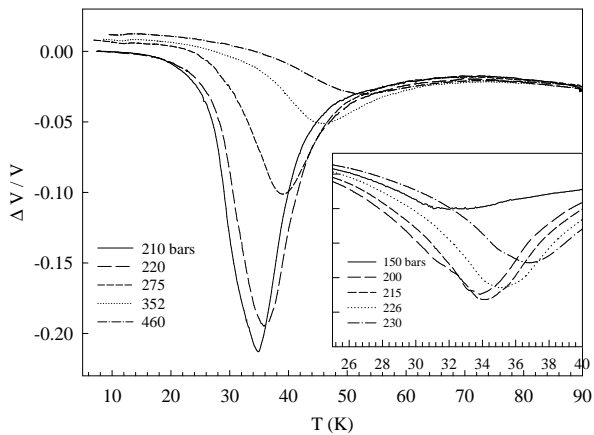}
\vspace{0cm} \caption{\label{}Temperature dependence of $\Delta
V/V$ at various pressures. The velocity variation is relative to
the value at 90 K. Inset: position and amplitude of the anomaly
below 230 bars.} \vspace{0cm}
\end{center}
\end{figure}

\begin{figure}[htbp]
\begin{center}
\includegraphics[width=8cm]{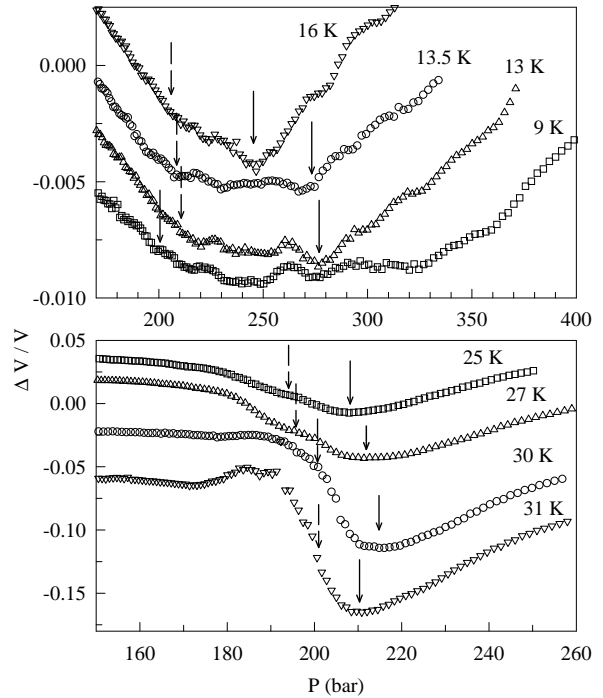}
\vspace{0cm} \caption{\label{}$\Delta V/V$ as a function of
increasing pressure at fixed temperatures. The velocity variation
is relative to the value at 150 bars. Curves have been shifted
vertically for a better view. Solid arrows signal the maximum
softening and dashed ones indicate the 200 bars feature.}
\vspace{0cm}
\end{center}
\end{figure}

The small anomaly in $\Delta V/V$ observed around 16 K for a
pressure of 270 bars (Fig.2) is due to the crossing of a quasi
vertical transition line in the $P-T$ phase diagram of Fig.1.
Since vertical lines are better identified from pressure sweeps,
we present, in figure 5, two series of increasing pressure sweeps
for different temperature ranges. In the top panel, we examine the
crossing of the superconducting phase boundary near 13 K. As the
pressure is increased above 170 bars all the $\Delta V/V$ curves
show a softening anomaly: at 9 K, the minimum of the anomaly
extends from 200 to 350 bars, while, at 13 K, the pressure range
is shortened to 200-275 bars, a trend which is still more
pronounced at 13.5 K. This clearly indicates that, at 13.5 K, we
have crossed the superconducting phase boundary. We defined the
small minimum around 275 bars (indicated by a solid arrow) as a
clear transition, while the 200 bars feature, corresponding to a
slope variation (dashed arrow), signals the onset of the
metastable SC phase. Above 13.5 K, the minimum progressively
deepens and moves to lower pressure (16 K curve in Fig.5). This
trend is more clearly observed at higher temperatures (lower
panel of Fig.5): the amplitude of the softening anomaly increases
and its position (solid arrows) shifts with temperature; there is
also a clear change of slope (200 bars feature indicated by
dashed arrows) on the low pressure side. The anomaly has maximum
amplitude ($\sim 20\%$) in the vicinity of the critical point
($P_0$, $T_0$). The 30 and 31 K curves of Fig.5 show a small
stiffening of the velocity around 190 bars, when the horizontal
crossover line, previously identified in the temperature sweeps,
is approached. Finally, for $T >$ 34 K (not shown), a softening
anomaly without the 200 bars feature is still obtained: its
amplitude decreases and it moves rapidly to high pressure along
the crossover line.

The two features indicated by arrows in the pressure sweeps have
been reported in the phase diagram of Fig.1. The right vertical
line (open squares) presents two parts. From 9 to 13.2 K, a
transition line centered at 275 bars separates two regions where
a metastable superconducting state is found. Then, this line,
starting from the same point (275 bars, 13.2 K), denoted ($P^*$,
$T^*$) in Lefebvre {\it et al.} paper \cite {Lefebvre}, has a
negative slope at low temperatures before going vertical as it
approaches the critical point ($P_0$, $T_0$) indicated by a gray
circle. Along this line, the softening of the velocity increases
dramatically to reach near 20$\%$ at the critical point, before it
joins with the crossover line. The left vertical line (open
triangles) defines, with the right one, a region where strong
hysteresis is found in the temperature sweeps (hysteresis is also
observed in pressure sweeps although it cannot be quantified).
This region begins at the SC-I boundary and, then, shrinks to zero
at the critical point where both lines merge and join with the
crossover lines. This hysteretic region coincides with the
previously observed first-order Mott transition
\cite{Lefebvre,Ito}.

We mention, finally, that, in the low pressure range, we could not
identify any clear anomaly related to a AF ordering transition.
This is highly surprising and unexpected since ultrasonic waves
couple generally quite easily with spin degrees of freedom
\cite{Trudeau}, although the absence of anomaly could signify a
very small coupling for this particular acoustic mode
(longitudinal polarized perpendicularly to the planes).

Beyond the confirmation of the $P-T$ diagram of a typical Mott MI
material, our ultrasonic data on the $\kappa$-Cl organic
conductor allowed the first detailed study of the critical point
region. The very large softening of the velocity at the critical
point corresponds to the predicted compressibility divergence of
the electronic degrees of freedom and validates then, the DMFT
picture of the Mott transition \cite{Kotliar,Onoda}. Two
crossover lines joining at the critical point were also obtained
from a similar, although smaller, compressibility anomaly. We
note the similarity with the two crossover lines identified in
the $Ni(S,Se)_2$ compounds \cite{Matsuura}. Most importantly, our
data reveal that the high pressure crossover line coincides with
the pseudogap features previously observed in magnetic
\cite{Mayaffre}, transport \cite{Shushko,Frikach} and elastic
\cite{Frikach} properties of organic compounds of the same family
having a higher chemical pressure. Since a Mott insulating phase
is present in underdoped high-$T_c$ compounds, this suggests that
the Mott critical point could play an important part in the
origin of the pseudogap in these materials. It should then be
interesting to investigate the pseudogap feature with an
ultrasonic measurement in high-$T_c$ compounds.

The authors thank C. Bourbonnais and A.-M. Tremblay for useful
discussions, suggestions and for the critical reading of the
manuscript. This work was supported by grants from the Fonds
Qu\'eb\'ecois de la Recherche sur la Nature et les Technologies
(FQRNT) and from the Natural Science and Engineering Research
Council of Canada (NSERC).

\bibliography{apssamp}

\begin{thebibliography}{99}
\bibitem{microwave}{We have measured the parallel and
perpendicular microwave resistivities at ambient pressure as a
function of temperature with a contactless microwave cavity
perturbation technique at 17 GHz. Below 50 K, the Mott-insulating
state is signaled by a rapid increase of the resistivity; we
observed, however, a change in the rate of increase (dip in
$d\rho/dT$) around 34 $\pm$ 2 K just above the AF transition
\cite{Lefebvre}, which is signaled at 27 K on the resistivity by a
strictly linear temperature dependence. This dip in $d\rho/dT$ in
the insulating phase has its counterpart in the metallic phase
where a peak in $d\rho/dT$, associated to the crossover line, is
rather observed. This dip in $d\rho/dT$ at $T$ = 34 K and $P$ = 1
bar is indicated as a dotted hexagon on the phase diagram of
Fig.1. }
\end{thebibliography}

\end{document}